\documentclass[12pt,a4paper]{article}
\usepackage[colorlinks=true, citecolor=blue, linkcolor=blue, urlcolor=blue, anchorcolor=blue, filecolor=blue, menucolor=blue, pdfborder={0 0 0}, breaklinks=true, linktocpage=true, pdfproducer=medialab, pdfa=true]{hyperref} 
\usepackage{cite} 
\usepackage{jheppubGenWG} 
\usepackage{graphicx}
\usepackage[titletoc]{appendix}
\usepackage{xspace}
\usepackage[utf8]{inputenc}

\makeatletter
\g@addto@macro{\UrlBreaks}{\do\-}
\g@addto@macro{\UrlBreaks}{\do\#}
\makeatother
\newcommand{\doi}[1]{\href{https://doi.org/#1}{doi:#1}}
\newcommand{\arxiv}[1]{\href{https://arxiv.org/abs/#1}{arXiv:#1}}

\usepackage{lineno}

\newcommand{\linenoDOC}{}

\title{\vspace*{-0.1cm} 
HL-LHC Computing Review Stage-2, \\
Common Software Projects:\\
Event Generators
}

\author[1]{The HSF Physics Event Generator WG\\Efe~Yazgan~(editor)}
\author[2]{Josh~McFayden~(editor)}
\author[3]{Andrea~Valassi~(editor)}
\author[4]{Simone~Amoroso}
\author[5]{Enrico~Bothmann}
\author[6]{Andy~Buckley}
\author[7]{John Campbell}
\author[8,9]{Gurpreet~Singh~Chahal}
\author[10]{Taylor~Childers}
\author[3]{Gloria~Corti}
\author[11]{Rikkert~Frederix}
\author[12]{Stefano~Frixione}
\author[13]{Francesco~Giuli}
\author[4]{Alexander~Grohsjean}
\author[7]{Stefan~Hoeche}
\author[14,15]{Phil~Ilten}
\author[9]{Frank~Krauss}
\author[16]{Michal~Kreps}
\author[17]{David~Lange}
\author[11]{Leif~Lonnblad}
\author[18]{Zach~Marshall}
\author[19]{Olivier~Mattelaer}
\author[7]{Stephen~Mrenna}
\author[20]{Paolo~Nason}
\author[21]{Simon~Plaetzer}
\author[22]{Christian~Preuss}
\author[23]{Emanuele~Re}
\author[3]{Stefan~Roiser}
\author[9]{Marek~Schoenherr}
\author[5]{Steffen~Schumann}
\author[24]{Markus~Seidel}
\author[7]{Elizabeth~Sexton-Kennedy}
\author[25]{Frank~Siegert}
\author[26]{Andrzej~Siodmok}
\author[3]{Graeme~A.~Stewart}
\author[27]{Aravind~Thachayath~Sugunan}
\author[26]{Zbigniew~Was}
\affiliation[1]{National Taiwan University, Taipei, Taiwan}
\affiliation[2]{University of Sussex, Brighton, UK}
\affiliation[3]{CERN, Geneva, Switzerland}
\affiliation[4]{DESY, Hamburg, Germany}
\affiliation[5]{Institute for Theoretical Physics, Georg-August-University G\"{o}ttingen, Germany}
\affiliation[6]{University of Glasgow, UK}
\affiliation[7]{Fermi National Accelerator Laboratory, Batavia, USA}
\affiliation[8]{Imperial College, London, UK}
\affiliation[9]{Institute for Particle Physics Phenomenology (IPPP), Durham University, UK}
\affiliation[10]{Argonne National Laboratory, Lemont, USA}
\affiliation[11]{Lund University, Sweden}
\affiliation[12]{INFN Sezione di Genova, Italy}
\affiliation[13]{Universit\`a di Roma Tor Vergata and INFN Sezione di Roma 2, Italy}
\affiliation[14]{University of Birmingham, UK}
\affiliation[15]{University of Cincinnati, USA}
\affiliation[16]{University of Warwick, UK}
\affiliation[17]{Princeton University, USA}
\affiliation[18]{Lawrence Berkeley National Laboratory, USA}
\affiliation[19]{Universit\'e Catholique de Louvain, Belgium}
\affiliation[20]{Universit\`a degli Studi di Milano Bicocca, Italy} 
\affiliation[21]{University of Vienna, Austria}
\affiliation[22]{Monash University, Melbourne, Australia}
\affiliation[23]{Laboratoire d'Annecy-le-Vieux 
de Physique Th\'eorique, France} 
\affiliation[24]{University of Maryland, USA}
\affiliation[25]{Institut f\"{u}r Kern- und Teilchenphysik, TU Dresden, Germany}
\affiliation[26]{IFJ PAN and Jagellonian University, Krakow, Poland}
\affiliation[27]{Tata Institute of Fundamental Research-B, Mumbai, India}

\emailAdd{efe.yazgan@cern.ch}
\emailAdd{mcfayden@cern.ch}
\emailAdd{andrea.valassi@cern.ch}

\usepackage{datetime2} 

\date{Version 1.1 (30 September 2021)} 

\newcommand{\mgamc}{MG5aMC} 

\newcommand{\mytexttt}[1]{{#1}}

\newcommand{\powheg}{Powheg} 
\newcommand{\minnlops}{MiNNLO$_\mathrm{PS}$}

\begin{document}
\linenoDOC

\maketitle

\section{Overview and Executive Summary}
\label{sec:execsummary}

This paper 
has been prepared 
by the HEP Software Foundation (HSF) 
Physics Event Generator Working Group (WG), 
as an input to the second phase of the LHCC 
review~\cite{bib:lhccreview,bib:lhcc145,bib:CERN-LHCC-2021-014} of 
High-Luminosity LHC (HL-LHC) computing, 
which is due to take place 
in November 2021.
It complements 
previous documents prepared
by the WG
in the context of 
the first phase of
the LHCC review in 2020, 
including 
in particular
the WG paper
on the specific challenges in
Monte Carlo event generator software for HL-LHC,
which has since been updated and  published~\cite{bib:gencsbs},
and which we are also submitting to the November 2021 review as an integral part of our contribution. 

Physics event generators 
are essential software components 
of the data processing and analysis chain 
of the LHC experiments,
and large consumers 
of their computing resources.
Because of the wide range 
of complex physics processes
which must be simulated,
the landscape of generator software 
is extremely varied,
including many 
packages developed by different teams.
A structured and comprehensive
review of the computational problems
posed by generators in the HL-LHC era,
including a description 
of technical challenges such as multi-threading or memory footprint,
and a quantitative estimate
of different types of 
computational inefficiencies
that can be addressed by dedicated R\&D
on both CPUs and GPUs,
has already been presented 
in the published WG paper~\cite{bib:gencsbs}
and we will not repeat it here.
In this new document,
we mainly go through
the generator packages
that are most relevant 
to the HL-LHC experimental program,
based on input 
collected in the WG in Q2 2021
through a series of dedicated 
meetings~\cite{bib:wgreview,bib:wgsherpa,bib:wgevtgen,bib:wgpythia,bib:wgmg5amc,bib:wgherwig,bib:wgpowheg,bib:wglhapdf,bib:hsfminutes}
with the teams
who are ensuring the development and maintenance
of these packages.
For each of these generators,
we address
the specific questions that
we have received as part of the charge
for the review~\cite{bib:wgreview},
while also discussing progress and plans on 
the issues that we had previously 
identified 
as high-priority for the WG.

We would like to start by emphasizing that the event generator area is not a single project (unlike ROOT~\cite{bib:root} or GEANT4~\cite{bib:geant4}), 
but consists of a variety of independent research and software projects~that~are managed independently by different teams, which are largely made up of theoretical physicists distributed across several institutes.
There are no centralized decisions on policies and strategies. In particular, the different generator and MC tool projects do not fall under the control 
of the same management structure as the LHC experimental programme,
also because in most cases these teams are not directly funded for that goal.
That said, there is clearly a (at least partial) convergence of interests between the goals of the generator projects and those of the LHC experiments, and there are several 
mechanisms for communication between generator teams and experiments.
While each generator team has its own agenda, which is largely dictated by several factors relevant to theorists' careers such as the publication of new theoretical research, these agendas are also heavily influenced by the needs of experimental analyses.
Many new developments in MC generator software follow specific requirements from the needs of the experimental groups. The generator experts in the experiments (for example, members of the generator or modelling working groups of the experiments, or analyzers who have specific needs) may communicate their issues or needs either directly to MC authors, for instance privately or with presentations or discussions in conferences and workshops, or through people involved in structures like MCnet or (in recent years) the HSF physics generators working group. 
MC authors address this wide variety of issues or requests as much as possible, but this depends on the person-power and work load in the MC team. 
Some of the problems do not have immediate solutions and may actually require years of technical and/or theoretical developments.
Understanding the current collaboration model between generator teams and the experiments, and finding ways to improve the 
management and funding (and, to some extent, communication) mechanisms
which are needed for the mutual benefit and success of both parties, are 
one of the main challenges in this area to be discussed in the LHCC~review.

The structure of this paper is the following.
Section~\ref{sec:execsummary}
includes an overview of the paper
and of the questions we addressed
during the review process,
a general description of how the generator projects are internally managed and how they interact with the experiments, 
and an executive summary 
of our main findings.
Section~\ref{sec:funds}
gives more details 
about the important issue of funding and careers,
which is common to all generator packages.
Section~\ref{sec:packages}
goes through various generator packages individually,
based on the input
collected from the corresponding teams.
Section~\ref{sec:common}
gives a non-exhaustive list
of ideas for possible common projects
in the generator area,
including more than one generator team
and/or external partners
such as the experiments, 
HSF, CERN IT or CERN EP.
We finally give some conclusions
and an outlook for further follow-up 
in Section~\ref{sec:outlook}.

\subsection{Review questions}
Among others,
we address the following questions
for each generator:
\begin{itemize}
    \item Are there plans and funds in place to continue support through HL-LHC?
    \item What major physics updates are foreseen for HL-LHC? Examples: transition from NLO to NNLO 
    (next-to-leading order to next-to-next-to-leading order) 
    calculations for ME+PS (matrix element plus parton showers), improved shower models, higher jet multiplicities, etc.
    \item What major software updates are foreseen for HL-LHC? What will be the main bottlenecks? In particular, what are the current CPU performance bottlenecks and how are they being worked on? Are there mitigation strategies to reduce negative weight fractions? Are improvements foreseen in phase-space sampling and unweighting efficiency?
     \item What work is in progress to adapt the software to GPUs and heterogeneous architectures for HL-LHC?
     \item Are there issues which may complicate the use of the generator by the experiments in multi-threaded workflows and what is the progress in this area?
     \item Is there any work in progress to include Machine Learning (ML) tools as part of the generator workflow?
    \item Are there issues or areas of work where help from HSF or from the experiments may be needed?
    \item Are there generators/tools not listed below that are expected or wished to become heavily used by the experiments?
\end{itemize}

\subsection{Review findings}

The following is a high-level summary 
of our findings:

\begin{itemize}

\item Funding and career opportunities are problematic.
While 
the interpretation and sensitivity of HL-LHC results 
heavily depend on 
the development and maintenance of 
generator software packages,
their authors 
(most often theorists and phenomenologists)
are generally not directly funded
to work for the HL-LHC experimental program,
and often lack the career incentives
to provide the software and computing 
support and
improvements needed by the experiments.
While 
the number
of core contributors 
with permanent positions 
and their personal interest
in the experimental program
seems currently enough
to provide at least basic support
for most generator packages,
there is limited funding for post-docs 
and PhD students. 
Thus, the loss of know-how
remains an important risk,
and in some cases 
this has already started to happen.

\item Concerning the problem of events 
with negative weights,
work is in progress 
in both MadGraph5\_aMC@NLO (\mgamc\ henceforth) and Sherpa
on improved NLO matching prescriptions.
The recent work on the MC@NLO-$\Delta$ 
technique
seems promising,
but there is not yet a clear timeline 
for its use in a new production 
version of \mgamc,
also because its implementation requires a strict coordination
 among the \mgamc\ and Pythia8 authors. 
The Sherpa team is also working towards achieving
 similar goals. 
It is important to note, however, 
that improvements in precision calculations, 
such as new parton shower algorithms
(e.g Dire)
and NNLO in QCD matrix element calculations, 
are expected to come 
with new sources
and an increased level of negative weights. 
Therefore, despite the recent progress, 
negative weights will likely remain an issue that needs close attention.

\item Several teams expressed an interest
in porting some of their software 
to GPUs, either directly
or through Machine Learning frameworks.
One development that looks especially 
promising is the ongoing work
on reengineering 
the matrix element calculation in \mgamc,
to exploit not only GPUs
but also CPU vectorization.
This work is making rapid progress
and might hopefully result 
in a new production version 
of \mgamc\ in the next year,
also thanks to a collaboration
between theorists and software engineers
which was largely promoted
by the HSF generator WG.

\item 
phase-space sampling is an important ingredient for efficient unweighted event generation. There are some new ideas and studies to improve unweighting efficiency, such as Machine Learning, for example, neural networks in Sherpa. 
However, these improvements 
have not yet been carefully validated
or used for large scale productions by the experiments.

\item Many analyses at the LHC depend on reweighted MC events to do systematic uncertainty calculations. PDF variations, parton shower initial- and final-state radiation variations, ME renormalization and factorization scale variations can be handled through computing alternative weights for the same physical event, saving CPU cycles. However, this cannot be done in some cases, such as for parton shower starting scale and underlying event tune variations (including color reconnections and intrinsic transverse momentum). This results in extra load in MC production systems of the experiments because of the additional CPU cycles to produce each sample with different parameters. 
Better flexibility in (re-)deriving scale and PDF weights can be obtained by writing more information to the event record (at least in Sherpa), but the CPU and storage implications of this have not yet been assessed.

\item Since the previous stage of the LHCC review, and having been highlighted strongly in the published WG paper, 
the generator groups are performing more in-depth profiling of the 
CPU costs and possible computational inefficiencies 
of their software. 
By clearly identifying some performance bottlenecks, this has already allowed major improvements in several generators, such as helicity recycling in \mgamc, an optimised program flow for weight variation calculations in Sherpa, and a new interface to MCFM’s analytic one-loop amplitudes; more details for each of these are available in the corresponding generator subsections in Section~\ref{sec:packages}. 
In parallel, a more detailed study of the CPU costs of MC generation campaigns in the experiments is progressing, also in terms of CPU times normalised to the \mytexttt{HEP-SPEC06 (HS06)} benchmark.
Establishing better mechanisms to collect CPU costs and generation efficiencies (e.g. due to sampling, merging or filtering) from the production systems of experiments would allow for an easier comparison between different experiments and different generators.

\item 
Several teams noted
that a collaboration with 
experts in the HSF 
would be very useful and welcome
to help them in the
optimization or
reengineering of their software.
One group mentioned that a software review,
like that organised by the HSF 
for the GeantV project in the past~\cite{geantreview},
may eventually be an interesting option.
We also believe that 
redesigning and harmonising software interfaces,
to improve package modularity
and ease the integration of plugins
from different projects,
would be a useful development, a view that was shared by some of the generator groups.

\end{itemize}

\section{Funding for generator work}
\label{sec:funds}
Funding and career perspectives are a major common issue reported by many generator experts.
Those working in the development, optimisation or execution of generator software provide essential contributions to the success of the (HL-)LHC physics program and it is critical that they get the right recognition and motivation. However, theorists get recognition on published papers, and may not be motivated to work on software optimisations 
that do not have enough theoretical physics content
to advance their careers. Generator support tasks in the experiments may also not be enough to secure jobs or funding for experimentalists pursuing a career in research. It is still not clear for people who are doing work on generators or related tools whether it is really possible to have a dedicated career path from now throughout HL-LHC. This is slowly being acknowledged from funding bodies, but the community has low expectations especially because there is no sign that hiring policies will be modified to provide reasonable funding for generator work. The delay caused by the lack of funding for generator work has already started to cause a loss of know-how in some generator packages. 

In general, MC tools (see Section~\ref{sec:mctools}) are even less supported than generators. The work involved for these tools has a lot of technical parts but little physics content.
One needs to find a good model to support this type of work. Such tasks may be ideal for Research Software Engineers (RSEs), an approach which seems to work well in the UK for similar situations: 
the idea is that these are relatively small, self-contained, tasks which do not require a large number of dedicated FTEs over a long time, but could be dealt with for example by allocating 10\% of the time of one person from a pool of RSEs with the required technical competencies.
Additional support from CERN EP-SFT for MC tools, which was possible in the past, would be very welcome, 
for instance through the GENSER project that is already supporting common builds and distributions of generator packages for the LHC experiments.

It should also be noted that  the EU ITN MCnet~\cite{mcnet} network has been used extensively for more than a decade by many generator teams, especially for more optimisation-focused work, but this project was not renewed and expires in few months. This funding was particularly important for MC tools, and needs to be replaced with another source to ensure their continued support.

On a positive note, some recent funding initiatives have explicitly targeted work in the event generator area. This is the case for instance for the DOE funding of the High-Energy Physics Center for Computational Excellence (HEP-CCE)~\cite{hepcce} and of the STFC funding of the SWIFT-HEP~\cite{swifthep} project, both of which could be interesting models also for other countries to fund software evolution in HEP.

\section{Review of individual generator packages}
\label{sec:packages}

In this section,
we individually go through 
the generators
that are most relevant 
to the HL-LHC experimental program.
Each subsection is based on the input,
and in most cases on the text contribution,
from the team
ensuring the development and maintenance
of the corresponding package.

In particular, we were explicitly asked~\cite{bib:wgreview} by the reviewers to cover \mgamc, \powheg, Sherpa, Pythia8, EvtGen and HepMC.
We then added Herwig7, LHAPDF and other MC tools.
There are also other generator-related software packages that have not been discussed in this document, even if they already are widely used,~like FastJet~\cite{fastjet},
or may eventually become heavily used by the experiments,
like \mytexttt{MATRIX}~\cite{Grazzini:2017mhc} and \mytexttt{Geneva}~\cite{Alioli:2012fc,Alioli:2015toa} or dedicated resummation frameworks built in event generator spirit like CVolver~\cite{DeAngelis:2020rvq}. 
We have also not covered any dedicated heavy-ion generators, such as 
EPOS~\cite{epos} or HIJING~\cite{hijing}.
All of these are likely candidates for topical discussions in the upcoming meetings that we plan to organize in our WG.

\subsection{MadGraph5\_aMC@NLO}

MadGraph5\_aMC@NLO~\cite{Alwall:2014hca} (\mgamc) is a multipurpose 
hard-event generator that carries out automated calculations
of SM and BSM
processes at the leading- and next-to-leading order (henceforth LO  and NLO respectively) in QCD. 
It may deal with any theory
whose Lagrangian can be specified by the user through an UFO~\cite{Degrande:2011ua} model,
including EFTs.
Such computations may be either at fixed order, or matched and/or merged to parton shower MCs. Fixed-order LO and NLO results are also available (see in particular Ref.~\cite{Frederix:2018nkq}) in the 
combined perturbative expansions in the QCD ($\alpha_S$) and EWK ($\alpha$) couplings, namely for any matrix element that factorises the quantity $\alpha_S^n\alpha^m$, with $n+m$ a constant. As such, the program features a significant flexibility, with extensive user-definable run and configuration settings. In keeping with its scope of hard-event generator, \mgamc\ only produces parton-level events, and delegates parton showering and hadronisation to external tools such as Pythia8 and Herwig7.

The \mgamc\ collaboration is a composite team of theorists with different expertise. Most members hold permanent positions and one is in a software-focused role. Software development is shared semi-equally across several members of the team. But the collaboration are of the opinion, shared by other groups, that time to work on performance improvements, even by significant factors, is not well recognised by funding agencies.

The version control of \mgamc\ is done through Bazaar. Its webpage~\cite{mg5webpage} in launchpad has extensive information and support.  

\paragraph{\bf Support through HL-LHC.}
 There are no dedicated funds for HL-LHC. The effort comes through different funding schemes with typically a theoretical emphasis. MCnet funding was used in the past for more optimisation-focused work but this project was not renewed and expires in few months. 

\paragraph{\bf Major physics updates for HL-LHC.}
Physics improvements are foreseen especially for higher-order EWK corrections, in particular for the extension of fixed-order capabilities in a QCD+EWK mixed expansion to include their parton-shower matched/merged counterparts.

\paragraph{\bf Major software updates for HL-LHC and main bottlenecks.}
The main consumer of CPU cycles
is, as expected, the ME calculation both at LO and NLO. Many optimisations are already in place for the loop calculations, so the NLO real emission is the main bottleneck there. Recent improvements, such as helicity~recycling, have been implemented,
achieving speed-ups~\cite{Mattelaer:2021xdr} ranging from factors 2-3 for standard processes such as $W$+jets and $t\bar{t}$, up to four orders of magnitude for some VBF-like processes, but these still need to be ported to NLO. When the issues above will be
 addressed or mitigated, \mgamc\ authors believe that one of the most time-consuming
 operations will be the computation of the colour algebra in an exact~manner.

Negative weights are an issue at NLO as in many generators.
Processes such as $V$+jets can have a negative weight fraction as high as 15-30\% for configurations with multiple additional partons. However, the negative weight fraction can depend more on the process than on the number of additional partons, for example, the negative weight fraction in $t\bar{t}$ production is larger than $V$+jets for the same final state multiplicity. 
Significant work has recently been undertaken to improve this in the MC@NLO-$\Delta$ technique~\cite{Frederix:2020trv}, which builds upon the MC@NLO matching
formalism~\cite{Frixione:2002ik}. The improvements from MC@NLO-$\Delta$ vary significantly depending on the configuration being considered, but factors of 2-3 reduction in number of events required for a given statistical power can be obtained even for standard processes like $W$+jet and $t\bar{t}$. The final implementation of this requires 
a strict coordination among the \mgamc\ and Pythia8 authors.

There has been much effort on improving the LO phase-space integrator~\cite{Mattelaer:2021xdr}. This mainly addresses VBF/VBS processes, 
where 
the gain in efficiency is very large. Neutral network approaches to phase-space integration, in collaboration with the VegasFlow~\cite{bib:vegasflow} team, are also starting.

Significant effort is being invested 
in the reengineering of the matrix element calculation, targeting GPUs (using CUDA but also abstraction layers like Kokkos, Alpaka or SYCL) as well as CPU vectorization in C++,
in collaboration with members of the HSF Generators Working Group from CERN IT and other institutes. 
With respect to the current Fortran implementation of the ME calculation, speed-ups by factors of 3 to 4 on CPUs through C++ vectorization were observed~\cite{bib:mg4gpu} for a simple electroweak process, while the CUDA implementation on a full V100 GPU was observed to be 1000 times faster than Fortran on a single core of a CPU. An additional speedup by a factor $\sim$2 was observed in both cases when moving the ME calculation from double to single floating point precision. 
While this progress is very promising, many essential developments like the integration of the new ME calculation in the existing Fortran framework for phase-space sampling and unweighted event generation, the backport to code-generating code to allow the ME computation of any LHC process, and a more careful analysis of QCD-specific computations like color algebra, are all still work in progress.
In addition, this work is restricted to calculation of LO matrix elements, and extending this to NLO will also require significant effort, in part due to added complexities, such as loop diagrams, some of which may require quadruple precision.
Just like most of the other speed-ups described in this document also for other generators, it should be stressed that a realistic extrapolation to achievable speed-ups and improvements for the LHC experiment productions will only be possible after many additional careful studies of software integration and physics validation.

\paragraph{\bf Issues or areas of work where help from HSF or experiments may be needed.}
There is already a lot of effective help in the GPU project\cite{bib:mg4gpu}. 
More coordination with experiments, either via HSF or directly, would be very welcome, for example in 
defining a better release strategy and on improving the communication between the generator authors and the experiments. 
\mgamc\ authors have specifically pointed out~\cite{bib:wgmg5amc} 
that the feedback cycle between generator teams and experiments on software updates, although much improved in recent years, would benefit from being made more efficient.

\subsection{\powheg}

\powheg \cite{Nason:2004rx,Frixione:2007vw,Alioli:2010xd} is a framework that hosts the code to simulate many physics processes. 
The main focus of \powheg\ is matching of accurate fixed-order predictions with PS for SM processes, mostly for LHC physics. Several BSM applications also exist, but BSM is not the main focus so far. \powheg\ is publicly available at \url{http://powhegbox.mib.infn.it} and is open to all theorists that want to contribute. There are about 100 processes from contributions of about 100 physicists. 
There are some ``core'' developers of \powheg\ but it is not a strictly well-defined collaboration. 
\powheg\ has fully supported interfaces for OpenLoops \cite{Buccioni:2019sur}, GoSam \cite{Cullen:2014yla,Cullen:2011ac}, Madgraph4, and \mgamc. 
The version control of \powheg\ is done through SVN. Its webpage has extensive reports for bugfixes and revisions. 
It has two main releases: \powheg\ BOX V2 which is the main release containing almost all processes, and \powheg\ BOX RES which is the most recent release created to deal with processes with resonances in particular. 

\paragraph{\bf Support through HL-LHC.}
Although there are no funds specifically dedicated to the maintenance of the framework, support and development will continue through HL-LHC by physicists with permanent positions who are already working on~\powheg, as long as there is interest in the LHC community.

\paragraph{\bf Major physics updates for HL-LHC.}
Major directions for physics updates are NNLO QCD+PS (\minnlops~\cite{Monni:2019whf,Monni:2020nks}), NLO EWK+PS, and interplay with modern PS. All of these updates, in particular the interplay with modern PS codes, may require major rethinking, restructuring, or re-coding of core parts of the software. 
Although it is difficult to predict, there are some possible processes that could be addressed to increase precision, such as loop-induced processes, for which some results already exist, and extremely high-multiplicity processes. 

\paragraph{\bf Major software updates for HL-LHC and main bottlenecks.}
ATLAS, CMS and LHCb have raised issues relating to difficulties due to the nature of independent implementations of the various \powheg\ processes. This requires a lot of non-user-friendly by-hand work to get compilation to run smoothly in the experiments' software environments. An improved central infrastructure would greatly improve this situation.
Considering the person power, the core software is expected to remain Fortran based with modern \mytexttt{f90} structures. There is growing interest in using GPUs for some aspects which will be detailed later. 

\paragraph{\bf Issues or areas of work where help from HSF or experiments may be needed.}
The \powheg\ authors already find it very useful to have contacts in the experimental community, in particular to test new developments especially before public releases, and also to test large scale production. 
HSF could coordinate refactoring and reworking that could be useful as a common work. 
\mgamc\ authors and others in CERN IT have been working together to make \mgamc\ work with GPUs.
For \powheg, a specific example could be identified, and this could start a collaboration between HSF, CERN IT, \powheg\ core authors and \powheg\ contributors who provide the ME calculations.  

\sloppy

\paragraph{\bf ME+PS generator specific considerations.}
As indicated above, the major directions for physics updates are to increase the precision of the predictions with NNLO QCD+PS (\minnlops),
NLO EWK+PS, and interplay with modern PS (e.g. PanScales \cite{Dasgupta:2020fwr}). 
LHC experiments have already started using some of the NNLO simulations from \powheg\ for the available processes. 
Until now, \powheg\ dealt with CPU bottlenecks through reweighting (for example with multiple PDFs, one or two loop amplitudes, scale variations), or through parallelization. 
In reweighting, calls to CPU-intensive routines are minimized to avoid re-computing. In parallelization, only using multicore has been sufficient without any complicated arrangements. 
It was noted by the experiments that some aspects of the reweighting functionality could be improved, such as the extension to physics parameter reweighting, $m_W$, $m_t$ and couplings, and also standard scale and PDF weights in \minnlops.

Some initial attempts have been made using MPI and there are plans to explore using GPUs or Machine Learning inspired techniques, for example for efficient phase-space sampling and for generation of underlying Born level events. The latter is the typically the more computationally delicate aspect of the \powheg\ core algorithm. It requires large statistics, and multidimensional integrals.  
Machine learning is being considered as a way to provide efficiency improvements in this area, but
no work has been done in this direction so far.

Negative weight events do not seem to be a major issue with respect to other generators, except at NNLO. 
There are already many processes available at NNLO QCD and they have negative weight fractions ranging from about 10 to 30\% depending on the process, which are much higher than the negative weights at NLO QCD in \powheg. However, at least for Drell-Yan with \minnlops, the fraction of negative weights can be substantially reduced (to a few percent) through folding. Larger negative weight fractions at NNLO are mostly due to the difference in scales in \minnlops\ and complications with photon emissions. It may also be possible to implement a positive resampler~\cite{Andersen:2020sjs,Andersen2021}. 

\fussy

In \powheg, having a harmonised framework to benefit from automated tools,
even just for NLO, would be very useful especially for some processes, e.g. for $W+c$ which has large negative weights in \mgamc\ but doesn't exist in \powheg. 
The implementation of new processes in \powheg\ could be made simpler through improvements of the existing interface to \mgamc. In particular, it would be useful if the interface exposed single amplitude contributions from the matrix element calculation, which are available in \mgamc\ (where they are used by MadEvent), but not yet to external generators like \powheg.

\subsection{Sherpa}
Sherpa~\cite{Sherpa:2019gpd} is a general-purpose generator that simulates hard interactions, radiative corrections, multiparticle interactions, hadronization, and decays. For the hard interaction it can simulate events up to NNLO in QCD (for selected processes) and NLO QCD with one-loop EWK.  

\paragraph{\bf Support through HL-LHC.}
Several of the Sherpa team are in permanent positions, and furthermore post-docs and PhD students have regularly contributed to Sherpa development and support over the last decades. In the UK, Sherpa posts at IPPP are seen as part of IPPP's core mission and therefore are expected to be treated as core in future funding reviews.

\paragraph{\bf Major physics updates for HL-LHC.}
The major physics updates foreseen for HL-LHC are an NLO parton shower, NLO multi-jet merging in decays, resonance-aware matching and NLO subtraction, NNLO matching for important processes, a full SM parton shower and NLO EW matching, SHRiMPS (inclusive QCD scattering simulation), improved bottom- and charm-quark decays, a fully validated Standard Model as an Effective Field Theory (SMEFT) and improved heavy-flavour parton evolution and NLO matching schemes.

\paragraph{\bf Major software updates for HL-LHC and main bottlenecks.}
The major software updates foreseen for HL-LHC are restructuring of event generation framework for improved efficiency, separation into ME and PS generator, and GPU support in ME generation. For restructuring of the framework and porting of code to GPUs, help from HSF and experiments would be extremely valuable. 
Performance analysis of the code has been done with the 
Intel Vtune profiler, MAQAO, MUST, and with Valgrind. 
A large part of the time is required to compute all weight variations. When weight variations are ignored, 
a typical multijet process spends 96\% of its time in the 
ME step, including phase-space and matching.
Without variations, it takes 1~s per event, versus more than 10~s per event with variations. And when variations are computed, more than 80\% of the CPU time goes into LHAPDF.
A significant bottleneck that has recently been identified and fixed is that many potentially large numbers of systematic variations (mainly for PDF uncertainties) were being calculated before the unweighting procedure, so for many events they would not in the end be used. The restructuring gives a factor of 5 speedup for a representative $V$+jets setup similar to the one used by ATLAS;
the eventual speedup achievable by ATLAS in production, however, can only be assessed by the experiment through further tests in its own software framework and configuration.
At the cost of theoretical precision, some computationally expensive aspects of NLO event generation can be simplified. For example, in some cases, spin correlations and color connections in the NLO matching procedure could be ignored~\cite{Danziger:2715727}. This is optional and already available in Sherpa 2.8.8. Some inefficiencies in the clustering algorithm are also being addressed.
After these optimizations, there are still three bottlenecks: one loop ME, PS clustering, and maths library. To improve one loop ME calculations, a generic interface to the MCFM~\cite{Campbell:2019dru,Campbell:2021vlt} library has been constructed. This sped up the computation of one-loop MEs by at least a factor of 6-10 with respect to automated one-loop providers, but is available for selected processes only.
 Another improvement for computing performance is expected to come from better phase-space sampling, to improve unweighting efficiency.
 To this end, the Sherpa collaboration explores variations of multi-channel integration techniques as well as Machine Learning~\cite{Bothmann:2020ywa,Gao:2020vdv}.
 The usage of Neural Network surrogates for the full event weight has shown significant potential to accelerate unweighted event generation \cite{Danziger:2715727,Danziger2021}. 
 An ongoing effort to calculate many-gluon tree amplitudes on GPUs~\cite{Bothmann:2021nch} is being extended to the full SM at present. A further efficiency improvement will be the reduction of the fraction of negative weights in NLO QCD calculations. There has already been significant progress in this direction~\cite{Danziger:2715727}, and there are several other ideas left to investigate. NLO parton showers will induce additional negative weights, whose rate can not be predicted reliably at present.

\paragraph{\bf ME and PS specific considerations.}
In the experimental analyses, some uncertainties such as PDF or ME scale can be calculated using weights stored in the samples. This saves experiments generating many extra samples for systematic variations.
Sherpa provides standard scale and PDF variation weights along with parton shower renormalization and factorization scale variations and uncertainties on NLO EWK corrections.
For HL-LHC, due to the long timescales involved, it may be desirable to have the ability to change PDF sets some time after a sample has been generated to avoid re-producing samples for systematic uncertainty calculations and modeling studies. Sherpa has the ability to write-out all coefficients needed to do a posteriori arbitrary scale and PDF variations into the HepMC event record~\cite{sherpa_aposte_pdf}. 
This offers more flexibility than the on-the-fly reweighting methods. 
However, for a MEPS@NLO~\cite{Hoeche:2012yf,Gehrmann:2012yg} event of a non-trivial multiplicity, or a high-multiplicity MEPS@LO event, this can easily be $>$50 to $\sim$100 coefficients.
The preferred usage mode is a strategic choice to be considered by the experiments for future large~productions.

Unlike the scale and PDF variations mentioned above, there are some variations which, in all generators, require alternative samples as they cannot be derived through event weights (for example, parton shower starting scale, tune variations, colour reconnections or hadronisation models). However, an example of another variation typically needing alternative samples, the merging scale, is being developed as a weight variation in Sherpa~3.0.0.

NNLO simulations are computationally very expensive, and the method of choice is likely to sacrifice precision in a controlled manner. For example, at NLO, color and spin correlations are typically dropped in PS matching.
In some cases there is very little difference between the predictions from an NNLO calculation and from an NLO calculation with additional corrections, so one could push to detector simulation only with NLO and apply some corrections for NNLO a posteriori. We need to understand the cases when we think we really need NNLO at the fully differential level.

\subsection{Herwig7}

Herwig7~\cite{Bellm:2015jjp} is a general purpose particle physics event generator. It computes any observable at NLO precision in perturbation theory matched/merged to a PS. It has two main shower modules: angular ordered and dipole-type, both including PS uncertainty estimated. It includes cluster hadronization and Eikonal MPI models and colour reconnection. It also allows automated BSM simulations using UFO model files. 
The release of Herwig7 was a significant update with respect to the previous Herwig++ interface. On top of the previous angular-ordered shower there was the addition of a dipole shower as well as a new \mytexttt{Matchbox} tool~\cite{Platzer:2011bc} allowing an interface to a large range of ME and loop providers and libraries, such as MadGraph and OpenLoops. In addition, there has been work on improving the logarithmic accuracy and including reweighting functionality for systematic uncertainties in the parton shower, and significant development on the minimum bias and underlying event simulation.

\paragraph{\bf Support through HL-LHC.}
The Herwig7 team is comprised of a core of permanent staff and a semi-regular cycle of post-docs and PhD students.
The plans and funds depend on institution and strategic goals.
Funding is a serious issue. Any funding for generator development must be strongly linked to a very strong physics case. In some cases this may not be what the experiments need. To fund the bread and butter of what the experiments need, one would need different specific funding.

\paragraph{\bf Major physics updates for HL-LHC.}
The major physics improvements foreseen are to accommodate higher-order EWK corrections and matched NNLO in QCD predictions. A full implementation of NNLO+PS matching 
requires developments of physics understanding in, and therefore changes to, the parton shower~algorithms.

An interface to string hadronisation model is under development. This would complement the existing cluster hadronisation model and allow for a much more robust estimate of the systematic uncertainties associated with these algorithms, which are projected to be dominant in some important physics goals at HL-LHC. 
 
\paragraph{\bf Major software updates for HL-LHC and main bottlenecks.}
To accommodate 
many physics improvements, such as those in the understanding of the shower,
a refactoring/rebuilding of some parts of the \mytexttt{ThePEG} framework, upon which Herwig7 is built, is needed. 

Negative weight fractions in Herwig7 for NLO processes have been prohibitively high for significant adoption in the experiments. Developments are in progress to improve these negative weight fractions, such as re-ordering contributions and resampling algorithms. This work also relies on addressing the structural problems in ThePEG previously mentioned. 
KrkNLO~\cite{Jadach2017}, an alternative matching scheme that can be used in Herwig7, has no negative weights, but it is not automated and is limited to only two processes for the moment. In addition, it cannot be used for merging and its systematics are not yet understood to the same level as in the default Herwig7 matching~prescription.

At this point, there is no work in progress to adapt the software of Herwig7 to GPUs and heterogeneous architectures. Things like showering are clearly a bottleneck for single instruction, multiple data (SIMD) processing. However, the ME calculation is vectorizable and Matchbox actually already heavily relies on interfacing to \mgamc\ for MEs, which will allow porting to GPUs. 

There is no work in progress to include ML tools as part of the generator workflow at this point.

\subsection{Pythia8}

Pythia8~\cite{Sjostrand:2007gs,Sjostrand:2014zea} is an event generator for high-energy particle collisions.  It contains physics models for hard interactions, radiative corrections, multi-parton interactions, hadronization, decays, beyond the standard physics, \mbox{$\gamma$-beams} and collisions, heavy-ion (HI) collisions, and other phenomenon.   A hallmark of the Pythia8 collaboration is providing good documentation, prompt responses to user questions and problems, and a rapid response to coding errors.  It has no dependencies on external packages like \mytexttt{ThePEG} or ROOT, is easily extendible with a UserHooks class, and has interfaces to many external packages and inputs, such as the (S)LHA interfaces.  The Pythia8 Collaboration (\url{https://pythia.org}) avoids duplicating efforts of other groups unless it is necessitated by physics or computing. 

\paragraph{\bf Support through HL-LHC.}
Pythia8 is an international and multi-institutional collaboration with no formal constitution. 
Like the other projects discussed in this report, the Pythia team has a few senior researchers, some junior researchers, and a rotation of graduate and post-doctoral students.
As such, the continued development of Pythia8 throughout the HL-LHC era depends on funding agencies and laboratory priorities, though these are not expected to change. The biggest challenge to the collaboration is finding permanent positions for junior people so that they can continue their valuable contributions. 

\paragraph{\bf Major physics updates for HL-LHC.}
Major physics updates foreseen for HL-LHC are NLO parton showers, resonance-aware matching and NLO subtraction, fully differential NNLO matching for important processes, full SM parton shower and NLO EW matching, automated ME-corrections interfacing with ME generators, forced hadronization and rare hadron production, and QED showers that also work for decays. 

\paragraph{\bf Major software updates for HL-LHC and main bottlenecks.}
In general, Pythia8 is not a computing bottleneck.   As a result, there has not been extensive or systematic profiling nor dedicated attempts to optimize the code.   Performance improvements could be expected in the time and memory usage in some merging configurations, parton shower variations,  multi-threading, and heavy-ion~collisions. 

Few of the proposed physics updates are trivial, and they will require some large code additions and a thorough validation process. VINCIA~\cite{Fischer:2016vfv} and DIRE~\cite{Fischer:2017yja,Hoche:2017hno,Hoche:2017iem,Dulat:2018vuy} do not have a large user-base at this point, but are now integrated directly to the core code, which was already a major update. 
Ref.~\cite{Brooks:2021hnp} presents the implementation of CKKW-L~\cite{bib:ckkwl} merging with VINCIA sector showers. This approach mitigates the bottlenecks in the standard multi-jet merging implementations. Ref.~\cite{Brooks:2021hnp} shows that the time spent constructing sector shower histories has an approximately linear scaling behavior with increasing number of hard jets and the parton-level event generation and the total memory usage do not increase. With the merging scheme in Ref.~\cite{Brooks:2021hnp}, it is already possible to have more than 9 hard jets from the shower. Using this approach, the remaining main bottleneck in simulating merged configurations would be on the fixed-order calculations. 

To estimate uncertainties related to the simulation of the parton shower, many CMS and some ATLAS simulations use the event-by-event parton shower variations available in Pythia8 through event weights. 
In the CMSSW~\cite{cmssw} environment, 
parton shower variations increase the run time by a factor of $\sim$3-4 on average. But here, since one run corresponds to many independent runs, there is a net gain in CPU time, memory and storage. 
However, it may still be necessary to find methods to improve performance. 

Pythia8 is "thread friendly" in the sense that multiple instances of the Pythia8 class can be used concurrently on different threads.   Of course, this is still dependent on the thread-safety of any external packages that are interfaced to Pythia8 at run time. 
For example, EvtGen is known to be \textit{not} thread safe.
CMS has made some simple tests, demonstrating than the use of multiple threads increases the speed of the simulation.  However, serializing and de-serializing data products takes considerable time, lessening somewhat the impact.
There is still significant room for improved MP/GPU/concurrency. 

It is not entirely clear how to leverage GPUs and GPU-based HPCs.   One option, which would still require some developments, is to have Pythia8 delegate matrix element calculation to the GPUs, for instance either by external calls of the \mgamc\ ME API or by reading HDF5 events.

ALICE, CMS and LHCb use the Pythia8 heavy-ion option ``Angantyr''~\cite{Bierlich:2018xfw}. 
Simulating HI collisions takes much longer compared to pp collisions. 
Simulating PbPb collisions (at $\sqrt{s_{NN}}=5.02$ TeV) is 
two to five orders of magnitude 
slower than simulating pp collisions depending on the configuration. For example,  hadronic re-scattering in HI collisions is a source of significant slowdown. 
More efficient reading of LHE files could also be investigated. 

\paragraph{\bf ME+PS generator specific considerations.}
A major use case of Pythia8 is in conjunction with external ME calculations that are matched or merged with the parton shower.
To understand better the current CPU performance, Pythia8 needs more dedicated profiling.  The acceptance efficiency is something that could be improved.
The veto method is currently delayed until late in the algorithm, which allows Pythia8 to calculate the no-branching probability and to conserve unitarity at a loss of CPU efficiency, but the extent of this inefficiency is currently not known quantitatively. 
Little is known about the large scale performance of VINCIA and DIRE. 
What \textit{is} known is that the VINCIA shower is an order of magnitude slower that the standard shower.   Promising areas of improvement are know, but no one is currently working on this. 
NNLO matching in Pythia8 is possible with DIRE and being worked on with VINCIA. 
At this point, no issues are foreseen for this, but it should be investigated more thoroughly.
Currently, Pythia8 does not offer any actual functionality for NNLO+PS, neither in DIRE nor in VINCIA.

For all updates, the main bottleneck is finding time and people. Help from HSF and experiments may be needed in restructuring of code, porting of code and algorithms to GPUs (if there is a use case), profiling for things people actually use for the HL-LHC era, and HDF5 usage and exploitation. Pythia8 would benefit from HSF support on issues that increase the common good in the generator world. Basic profiling for relevant use cases would be beneficial for improving Pythia8 computing performance. 
VINCIA and DIRE are recent major updates that needs to be tuned to data and to be test-driven at a larger scale by experiments to provide feedback both on physics and computing performance. 
One foreseeable issue with using DIRE is the significant negative weight fractions, especially when DIRE is used with NLO ME that already has negative weights. VINCIA does not have negative weights and tries to avoid them as far as possible in future developments. 
The Pythia8 authors expressed an interest in embarking on a GPU port through collaborations with HSF members and the experiments, similar to the ongoing work with MG5aMC.

\subsection{EvtGen}

EvtGen~\cite{Lange:2001uf} is a package specialised for heavy-flavour hadron decays. It consists of about 130 decay models which implement the specific dynamics of various decays. 
EvtGen maintains a detailed decay table with a large number of processes. Known decays do not add up to 100\% branching fraction. What is missing is filled up by generating quark configurations and passing those to Pythia8 for fragmentation. The $\tau$ decays are done through Tauola~\cite{Davidson:2010rw}, and Photos~\cite{Barberio:1993qi} is used for radiative corrections. 
The code has been largely stable over past 10 years with most of the changes due to added models. Some modernisation and cleanup was done rather~recently. 

\paragraph{\bf Support through HL-LHC.}
The core EvtGen team has 4 members who have a fraction of their time funded for EvtGen through an UK STFC grant.
Part of the funding is attached to LHCb as part of the responsibility (for general maintenance) within the experiment. One of the team members is funded by Warwick university as part of a collaboration with Monash. Beyond the core team, others contribute by writing new decay models driven by the needs of measurements. 
There is some funding for a computer engineer to work on code redesign for multithreading. 

\paragraph{\bf Major physics and software updates for HL-LHC and main bottlenecks.}
No updates for physics is foreseen in the near future.
Currently, the team is working on some code consolidation to unify coding style and for \mytexttt{C++} modernization. 
There is a plan to decrease code duplication within the decay models and to have an updated documentation to be in Doxygen, as well as a paper or a manual. 
There seem to be difficulties with the multithreaded environments that experiments are moving to, due to issues with thread safety, but the plan is to allow event level multithreading. In this context, event means the particle to be decayed through the full decay chain. 
The main bottlenecks are Tauola and Photos, neither of which allows multithreading. There are also structural limitations in EvtGen for running multithreaded mode, such as a global instance of random number generators.
Structural changes can be done, but this requires a major re-write of the code. The issues in Tauola and Photos could be fixed with more time and help. It is also possible that, for the dependencies, Tauola might be replaced by Pythia8 and the interface to Photos may be improved or instead an alternative such as SOPHTY~\cite{Hamilton:2006xz} may be used. Photos is used in practically in every decay in the usual use cases. It is seen from profiling that significant CPU is spent in Photos itself and in EvtGen the conversion to and from HepMC is significant. 
Pythia8 provides $\tau$ decays using the helicity formalism. EvtGen already depends on Pythia8 for other reasons, but that functionality could potentially be used for $\tau$ decays. The main interface is being finalized. 
Unlike Tauola, Pythia8 provides amplitude for particular spin states and sums over spin states are done by EvtGen. One challenge is passing the spin states from EvtGen to the tau decayer. This is being finalized as well. The decay table is one of the things that is difficult to update because of unobserved (i.e., explicitly listed) decays and consistency of particle properties between generators. It is important to agree with other generators about particle properties. 
In LHCb, the new version of the simulation under development uses a single EvtGen instance which is locked as needed for multithreading. LHCb, by default, runs generator and simulation in the same job. This is not the same in ATLAS or CMS in all cases. 

\subsection{MC Tools: LHAPDF, HepMC, Rivet, and others}
\label{sec:mctools}

Tools to support and build on MCs are an important part of the ecosystem of collider phenomenology. They are socially distinct from typical event-generator development, the teams working on them being as much or more rooted in the experimental than theory community. Sustainability is clearly an issue since most of this work is not funded, particularly with the end of more than a decade of funding through the EU ITN MCnet network. MCnet has provided funding in particular for MC-development PhD students and for theory and experiment PhD students to contribute to generator-project initiatives, and for travel and meetings to enable collaboration between developer and user groups. It is undoubtedly the case that development of shower+hadronisation MC generator features, and in particular the tools that connect them to experiment and phenomenology, will be curtailed by the end of MCnet-supported activities. This funding needs to be replaced with another source to ensure continued support and development of these essential tools.

\paragraph{\bf LHAPDF.}

LHAPDF~\cite{Buckley:2014ana,Andersen:2014efa} is a library for PDF-value access, and associated metadata including $\alpha_s$ calculations consistent with the PDF fit. LHAPDF was redeveloped from v5 to~v6 in 2012/13 as a purely interpolation-based tool rather than recalculating DGLAP evolutions as with some older PDF sets in v5 and earlier. It defines the standard data format for all PDF-set data, is the standard PDF interface for LHC-experiment MC production, and is MPI-friendly.

However, its CPU cost is unnecessarily high: profiling of Sherpa+LHAPDF in experiment-like NLO $V$+jets setups has shown that the main CPU consumers are calls to LHAPDF and transcendental functions, e.g.~\mytexttt{log}. Regarding the latter, there could be a 10\% gain in LHAPDF performance simply from runtime use of the Intel math library; this is already in use by ATLAS, with other benefits throughout the software chain.
In 2020, \mytexttt{LHAPDF 6.3} introduced thread-safe caching as a response to these studies, in particular to benefit from thousands of identical $(x,Q^2)$ phase-space-point queries across the set of parton flavours, or large PDF-fit ``error sets''. Efficient use of this caching, however, depends on the generator codes using a commensurate call strategy: repeated calls to the same phase-space point are often not batched together, but separated by many other PDF evaluations, forcing cache misses.

A final MCnet short-term studentship project has focused further on optimisations such as data restructuring and precomputation of repeatedly used functional derivatives, and will prototype a GPU-based interpolation feature for use with GPU-based matrix-element sampling. A Google Summer of Code (GSoC) project on implementation of higher-order interpolators further revealed improvements to the interpolation heuristic, and clarified that for N3LO calculations the precision of the interpolation may in fact be less important than its global smoothness: this motivates future development of alternative interpolator strategies which prioritise higher-derivative continuity over either precision or CPU efficiency. A UK SWIFT-HEP part-time research assistant on LHAPDF and corresponding MC efficiency will extend this work starting from October 2021. 
CERN EP-SFT provides 
part-time staff support for adding new PDF data to the EOS-based PDF-set repository, previously hosted at Durham IPPP.

In summary, \mytexttt{LHAPDF} and HepMC are crucial to MC production in experiments. There is a huge PDF-interpolation cost, which can be largely addressed by concerted efforts to restructure generators to make more efficient PDF calls; trivially, by lazily evaluating error sets in ME sampling, and less trivially by reorganising call orders to make better use of caching, either in the generator or in LHAPDF. Further improvements to the interpolation library, for focuses on efficiency, precision, and smoothness are also required, as are extensions to support generalised PDFs (in particular resolved-photon PDFs for EIC and FCC-eh, and GPDFs and DPDFs for small-$x$ hadron-collider physics). All these developments, technical and physics-oriented, require explicit community support.

\paragraph{\bf HepMC, LHE and other MC event formats.}

HepMC~\cite{Dobbs:2001ck} is the standard library for event-graph representation. HepMC v1 had undergone problematic forking between the CLHEP project and the LHC experiments, which was resolved with v2 before the start of LHC data-taking. However, this version retained other structural problems, including an unmaintainable API, poorly organised code, ROOT I/O unfriendliness, logical inconsistencies in e.g. const-correctness, and limitations in handling of heavy-ion and other requirements from specialised areas of collider physics. HepMC3~\cite{Buckley:2019xhk} is a major ground-up rewrite, addressing these issues and making the library more extensible. CERN EP-SFT helped with the v3 rewrite, project management and some funding, 
until the rewrite was~finished.

There is still much work needed to optimise the efficiency of the graph algorithms used to write and explore HepMC event-graphs, and technical \mytexttt{C++} optimisation of the data structures. Event copying, graph-walking, and filtering/rewriting are CPU-expensive, which can likely be reduced by expert technical developer time. New features such as the ``attributes'' have not been optimised for I/O or CPU efficiency, and are similarly amenable to improvements. Usability is another area ripe for development: more expressive search and filtering functionality in the API, and provision of a set of command-line tools for event-file splitting, merging, and filtering would enable more efficient workflows for experiment and phenomenology users alike. This character of ``utility development'' is unlikely to occur without paid, dedicated effort, as there is little academic reward for such contributions.

Smaller graphs which are physics subsets (size smaller by a factor of 10) could offer significant benefits to users both in storage and CPU time. A discussion on defining such subsets and ensuring their implementation in parton-shower event generations should take place, via the MCnet community, HSF, and/or the Les Houches~workshops.

In terms of I/O formats, HepMC3 supports (zipped) textual formats from v2 and v3, and ROOT. Read and write support for further formats can be easily added, if other formats are necessary (e.g.~from the heavy-ion community as it moves to increasingly use LHC standard tools in the EIC and elsewhere. Format improvements are possible, but must be undertaken judiciously to avoid breaking read-compatibility with previous versions: this can be handled explicitly in the custom formats, but schema evolution is more complicated with binary formats such as ROOT and HDF5.

Handling parton-level events is also important. The \mytexttt{LHE} format is textual and not fully standardized. It may be required to move to a more efficient format such as \mytexttt{HDF5}. 
\mytexttt{HDF5} support was added by the Sherpa and Pythia8 teams, but it is also not standardized. 
HepMC3 can embed \mytexttt{XML LHE}, however is it not clear how ME and showering and hadronization event generators (SHG) event-record should be linked for \mytexttt{HDF5}. None of these required developments are institutionally supported. This is fully a technical question that has no career incentive but it requires a good degree of software design and engineering experience. In Pythia 8.304, a new versioning scheme for HDF5 event samples exist, \mytexttt{LHAHDF5}~\cite{ref:lhahdf5}. This is based on the structure defined in Ref.~\cite{Hoche:2019flt}.

We need better graph algorithms, copying, manipulation tools and small event-graph options from MCs.

\paragraph{\bf Rivet.}

Rivet~\cite{Bierlich:2019rhm} is an analysis-preservation tool, used in MC validation, development, tuning, and BSM reinterpretation. 
Rivet v3 was released in June 2019, with the current version being 3.1.4. This latest release series notably provides automatic MC multi-weight handling for systematic uncertainty calculations, improved run-merging (that allows things like ratios to be calculated correctly without any assumptions), machinery for heavy-ion observables, and the ability to pass optional parameters to analyses.

Possible development areas are statistical-object handling, HepData integration, detector folding and smearing, systematic uncertainties and bootstrap, as well as technical developments like HDF5 formats to enable speed-up, and embeddability in adaptive-sampling physics tools, e.g. for MC-tuning and BSM limit-setting.

The code efficiency was profiled in GSoC~2020, and found to be limited~by~HepMC in I/O, event copying for unit standardisation, and event-graph traversal. Multi-weighted events provide opportunities to cache histogram operations, which currently occupy the next largest fraction of CPU time. This and thread safety require detailed technical work, as there are few simple opportunities for efficiency improvements.

Many suitable LHC experimental results are preserved using Rivet, with semi-active preservation programs in experiments. The official requirements in the experiments are not currently strong, but they exist and are getting better. The important role of Rivet in MC development, validation, and BSM fitting requires better standards-compliance in the routines submitted to Rivet. The experiment-published analysis coverage (mostly of ATLAS and CMS for now) is tracked in a Rivet webpage~\cite{rivet_webpage}, using publication metrics from \mytexttt{Inspire}. ALICE has initiated and strongly contributed to the RIVET for HI implementation. Their analysis contribution is ramping up and also non-LHC HI experiments have started to contribute.
If analysis preservation is not foreseen from the beginning of experimental measurements, it is more likely to encounter issues with Rivet integration or downstream usage.

There is also a critical issue that the analyses submitted with reference data are sometimes incompatible with that submitted to \mytexttt{HepData}. The untangling of this incompatible portion of the Rivet data is draining the Rivet resources. Rivet no longer accepts analyses that are not \mytexttt{HepData}-compatible, therefore, this won't be a problem in the long term.

Rivet development was greatly impacted by the Covid19 pandemic, with fewer patch releases than usual and no development toward the planned major updates. Much of the development effort needed is in areas involving complex statistical tools, e.g., to handle event weights. 

\paragraph{\bf YODA, Professor, MCUtils/HEPUtils, HepData.}

These are other, ``small'' but important, tools that are even less supported than those described above.
This is a work it is hard to get rewarded on, in particular it is difficult to write papers on these tools. 

\mytexttt{YODA} (\url{https://yoda.hepforge.org}) is the statistics library underpinning Rivet. 
It is lightweight and focused on the issues encountered in MC event-analysis and analysis reinterpretation,
and it is easy 
to integrate in multi-threaded applications because it has no global state.
It is being extended for generic-object binned containers, coherent many-dimensional histograms, correlations, and to have a new binary format. However, there is again no funding support or dedicated person-power.

\mytexttt{Professor}~\cite{Buckley:2009bj} is an MC-tuning and general parametrization and optimization tool used to make many of the main LHC MC-generator tunes. It has been rewritten as \mytexttt{Apprentice}~\cite{Krishnamoorthy:2021nwv} with added support for rational approximants, better handling of correlations, and other improvements. Again, it is unsupported, although it has started to be used in MC teams of LHC experiments at least of ATLAS and CMS. 

\mytexttt{MCUtils/HEPUtils} are general tools to help with MC analysis, including HepMC graph filtering and PDG~ID decoding. HepMC3 development may be able to integrate the search and graph-reduction algorithm into the HepMC tool suite.

\mytexttt{HepData} (\url{https://www.hepdata.net}) is another important service that is not strictly software run by the experiments, 
but that holds an important place to store and preserve data used as input to tools such as Rivet. 
We note that continued support of HepData is also essential for the event generators ecosystem. Operation-wise, it is supported but there is little development. LHC data-reinterpretation studies make heavy use of \mytexttt{HepData} and have demands which would be best solved by feature development in the database system, particularly around provision of correlation information, semantic metadata, and preservation of theory predictions.

\section{Opportunities for common projects}
\label{sec:common}

\paragraph{\bf Proposal for modular MC event generator framework.}
There was appetite from a number, but not all, of the generator groups to create a modular framework with well-defined software interfaces where different teams can plug in different ME generators or different parton showers, even different shower algorithms and different hadronisation models, etc. 

There are several places where such an interface would not be a giant leap from what is already in place. For example: Sherpa is able to load Pythia8 libraries for hadronisation and hadron decays and choose between BlackHat~\cite{blackhat2008,blackhat2014}, MadLoop~\cite{madloop}, MCFM, OpenLoops, and Recola~\cite{Actis:2012qn,Actis:2016mpe,Denner:2017vms,Denner:2017wsf} for the loop ME provider; Herwig7 is working on a string hadronisation model based upon an old version of Pythia; and \powheg\ already has an interface to \mgamc. Pythia8 has a new generic ME interface framework, currently able to load C++ ME libraries generated with MG5aMC, but currently being extended by a runtime interface to the Comix ME generator in Sherpa and 
the new C++ interface~\cite{Campbell:2021vlt} 
to MCFM.
This was also
the initial idea behind \mytexttt{ThePEG}, which Herwig7 continues to use, but as noted  significant redevelopment would be needed to make this more widely~useful.

Such a framework could be very beneficial from both a physics and software point of view. For physics studies it would make the estimation of uncertainties in a factorised way much simpler. From a software point of view it would allow greater possibilities to improve and/or port code when a) the workflow is clearer to outsiders and b) there is a common structure. In order for this to happen, HSF could help identify and agree on some interfaces, but such a project would likely only be successful with support from the vast majority of generator groups.

More generally, taking the time to redesign some internal and external software interfaces of the generator packages would also be a good opportunity to improve the readiness of the code for modern software paradigms and hardware architectures. 
During the ongoing 
reengineering of the ME calculation in MG5aMC~\cite{bib:mg4gpu},
for instance,
it has become clear that 
the redefinition of some interfaces to deal with multiple events in parallel (rather than one event at a time),
as well~as~a~clearer encapsulation of the inputs and outputs of each software block (avoiding possible pitfalls from Fortran common blocks),
are required in order to exploit event-level data parallelism on GPUs and vector CPUs. This would also simplify the implementation of task parallelism via multi-threaded~approaches.

\paragraph{\bf Help in the effort of 
porting MC generators to GPUs.}
Experiments have GPU resources where they can test new MC workflows and validate them with running large scale validations, and can identify technical difficulties.
HSF can help MC authors and experiments identifying, refactoring, profiling and reworking that could be useful as a common work to make ME calculations with GPUs. 
HSF, \mgamc\ authors, and CERN IT are working together on GPUs for \mgamc. This could be generalized to include other MC generators and HSF can also help in the coordination with experiments as well.

\paragraph{\bf Basic profiling and improving computing performance.}
Generators may benefit from HSF support on issues that increase the common good in the generator world. Basic profiling for relevant use cases would be beneficial for improving computing performance. Thanks to a large effort from the generator teams in both experiments and the WG, a lot of insight into the settings used to support each experiment’s physics program was gained. A more detailed study of the different strategies is ongoing, in particular by analysing individually the CPU costs using standard benchmark suites such as \mytexttt{HEP-SPEC06 (HS06)}.
The work on establishing better mechanisms to calculate CPU costs using standard benchmarks from the production systems of experiments is ongoing. This will allow for more fair and easier comparison between different experiments and help identify bottlenecks in production. For profiling, another important development would be to have the MC generators report on sampling, merging, filtering and other relevant efficiencies. 

For performance profiling and comparisons,
it would be useful to have a standalone reproducible common setup for ATLAS and CMS. This is being worked on for the $t\bar{t}$ final state by the LHC Top WG, in collaboration with the HSF Physics Generators WG. 

\section{Conclusions and Outlook}
\label{sec:outlook}

In this note,
we have discussed the progress and plans
for some of the main event generators and tools
used by the LHC experiments,
taking the perspective of considering them
as common software projects that are essential to the success of the HL-LHC programme.
This complements the recent published paper~\cite{bib:gencsbs},
also prepared in the context of the LHCC review of HL-LHC computing,
where our WG had provided a comprehensive discussion of the challenges in event generator software,
and which we are also submitting to the November 2021 review as an integral part of our contribution. 

The ongoing and planned developments and the main issues for the various generators have been presented. It is found that funding and career opportunities are still problematic. There is progress in the problematic area of events with negative weights, however new higher order predictions are expected to yield an increased level of negative weights: therefore, negative weights will continue to remain an issue and require even more developments. There is rapid progress in the porting of matrix element calculations in \mgamc\ to GPUs thanks to a collaboration between MC authors and software engineers, which has been promoted by the HSF generator WG. There are some developments in improving phase-space sampling for efficient unweighted event generation and more are expected to come. The generator and tools authors are now paying much more attention to CPU efficiency and major gains have already been made to improve their software. Possible improvements are also discussed in event reweighting, 
which would reduce the load on the MC production systems of the experiments. Improvements in benchmarking and profiling in the experiments have also been discussed. Some opportunities for common projects have been highlighted. 
Other known sources of inefficiencies in the experiment workflows, such as those due to complicated filtering needs, still need to be worked on to identify possible ways to mitigate them with new developments in generators and~tools.
The increase in computational costs for MC generators in the HL-LHC era due to the need for more precise NNLO calculations, which we started to discuss in our WG~\cite{bib:gencsbs,bib:wggrazzini}, also needs to be carefully assessed.

Finally, we note that defining a specific roadmap for R\&D in the generator area for the next few years is a very complex and long iterative process, for which we believe that we are now only at the first step. As we already discussed extensively~in~our published WG paper, this is particularly difficult as this work mainly needs to~be~done by several 
independent teams of theorists, who are in most cases not directly funded to contribute to the HL-LHC experimental programme.
An important next step is to communicate with the generator/tools teams to obtain more specific prioritisation of the different development streams, the person effort required to achieve them and the performance gains expected. Following this, further iterations with the experiments will be needed to try to 
ensure that these aims are aligned with their priorities.

\end{document}